\documentstyle[aps,prl,amsmath,amssymb]{revtex}
\draft
\begin{document}
 \title{Small-q phonon-mediated superconductivity in 
organic $\kappa$-BEDT-TTF compounds}
\author{
Georgios Varelogiannis}
\address{
Institute of Electronic Structure and Laser,
Foundation for Research and Technology Hellas\\
P.O. Box 1527, GR-71110
Heraklion Crete, Greece} 
\date{\today}
\twocolumn[\hsize\textwidth\columnwidth\hsize\csname@twocolumnfalse\endcsname
\maketitle
\begin{abstract}
We propose a mew picture for superconductivity
in $\kappa-(BEDT-TTF)_2X$ salts arguing that
{\it small-{\bf q} electron-phonon} scattering
dominates the pairing.
We reproduce the distinct
X-shaped d-wave gap
reported recently by magnetooptic measurements and we argue
that 
the {\it softness}
of the momentum structure of the gap and the
near degeneracy of s- and d-wave gap states
may be at the origin of
the experimental controversy about the
gap symmetry.
We show that a magnetic field applied parallel to
the planes may induce extended gapless-regions on the FS accounting for
the experimental signatures of a Fulde-Ferrel-Larkin-Ovchinikov 
(FFLO) state
and it may induce
{\it gap symmetry transitions} as well.

\end{abstract}

]
\narrowtext
%


The fascinating physics of organic metals and superconductors (SC)
continues to motivate intense  
invenstigations \cite{Book,LangReview,Annett1}. 
Particularly interesting are the quasi-two-dimensional organic SC    
based on the donor $BEDT-TTF$ 
(bisethylenedithio-tetrathiafulvalene).
The $BEDT-TTF$ molecules  
are packed in various motifs (labeled by Greek
letters) in layers which are separated by
insulating ion planes.
The $\kappa$ packed compounds
exhibit numerous similarities with
high-$T_c$ cuprates
\cite{McKenzie,Book,LangReview,Annett1}.
As verified by de Haas-van Alphen and Shubnikov-de Haas
measurements \cite{Wosnitza2,SingletonR}
their carriers are usually quasi-dispersionless perpendicular
to the planes exhibiting the behavior of
almost perfectly two-dimensional metals.
When associated to a monovalent ion
$X$ (like $I_3$ or $Cu(NCS)_2$ or $Cu[N(CN)_2]Br$ or
$SF_5CH_2CF_2SO_3$ etc.)
with stoichiometry $\kappa-(BEDT-TTF)_2X$
(usually abbreviated as $\kappa-(ET)_2X$) they show     
SC
at temperatures that can exceed
$10K$ 
\cite{Book,LangReview,Annett1}.
At least in
$\kappa-(ET)_2Cu[N(CN)_2]Br$, $\kappa-(ET)_2(SCN)_2$ and
$\kappa-(ET)_2I_3$, specific heat, 
NMR relaxation, thermal conducitivty and some penetration depth
measurements point clearly to the presence of gap-nodes \cite{ExpDwave}.
On the other hand, other sometimes similar 
experiments indicate instead
a fully gapped superconducting state \cite{ExpSwave}
and a stricking
experimental controversy persists \cite{Annett1}.

Recently, a millimeter-wave magneto-optical technique
allowed to measure the angular dependence of the gap
in $\kappa-(ET)_2Cu(NCS)_2$ \cite{Schrama}.
The SC gap exhibits a distinct X-shape pointing to 
an anisotropic d-wave structure with nodes
along the 
{\bf b} and {\bf c} directions \cite{Schrama}.
The reports      
of a d-wave gap                       
motivated 
theoretical investigations of spin fluctuations (SF)
mediated SC 
in $\kappa-(ET)_2X$ salts \cite{Moriya,SpinFl}.
The potential relevance of SF has been justified 
by the proximity of antiferromagnetic (AFM) phases
in the pressure-temperature phase diagram \cite{AFMexp}.

However,                  
there is substantial experimental evidence
suggesting
that phonons are crucial 
for the pairing in
$\kappa-(ET)_2X$ salts. 
{\it Isotope effect}
measurements on $\kappa-(ET)_2Cu(SCN)_2$
\cite{Kini} report direct evidence
for a phonon mechanism.                                    
Raman spectroscopy on
$\kappa-(ET)_2Cu [N(CN)_2]Br$ \cite{Pedron} also 
advocate a 
phonon mechanism.              
Inelastic neutron scattering studies report
SC-induced
frequency changes in the intermolecular phonon modes
of $\kappa-(ET)_2 Cu (NCS)_2$ \cite{Rietschel}  
suggesting their involvement in the pairing.
Moreover,
the relevance of
strong electron-phonon scattering is firmly established 
by the increase of the lattice conductivity
(the phonon contribution to the thermal conductivity)
at the SC transition \cite{Behnia}.
Thermal expansivity measurements \cite{Thermal}
confirm a strong electron-lattice coupling which
may naturally dominate the pairing.

In the present Letter we introduce a new picture
for SC in $\kappa-(ET)_2X$ salts.
We show that 
{\it small-{\bf q} phonon mediated SC}
reproduces accurately the
experimentally observed X-shaped {\it d-wave
gap} in $\kappa-(ET)_2Cu(NCS)_2$
\cite{Schrama}.
We emphasize the 
distinct qualitative behavior exhibited by our SC states   
which is
related to the {\it softness} of the momentum structure
of the gap and
the near degeneracy
of s- and d-wave gap symmetries which 
may be at the origin of the experimental 
controversies in $\kappa-(ET)_2X$'s. We study   
the effect of an in-plane magnetic field
on our SC state which is shown to account for
recent experimental signatures  
of inhomogeneous SC near the in-plane critical field 
and may plausibly induce s-d gap symmetry transitions

The $\kappa$-packing motif is illustrated in Fig. 1a
where each stick corresponds to a BEDT-TTF molecule.
There are two different types of pairs of closely packed
BEDT-TTF
molecules called {\it dimers} and a unit cell is constituted by 
two dimers, one of each type.
Because the 
intradimer hoping is more than twice
the interdimer one     
and the splitting between the bonding and andibonding
orbitals is about twice the intradimer hoping,
only the antibonding orbitals contribute to the
Fermi surface. This allows to   
consider a dimer as the effective basic structural unit  
making the so called 
{\it dimer model approximation} \cite{Caulfield,Fukuyama}
which leads to an effective
frustrated lattice model illustrated in Fig. 1b
and to an
extended Brillouin zone (BZ) scheme
(see Fig. 1c).
Shubnikov-de Haas measurements \cite{Caulfield}
confirmed the relevance of
the dimer approach in $\kappa-(ET)_2X$ salts which corresponds
to an electronic dispersion of the form 
$$
\xi_{\bf K}= 2t \bigl( \cos K_y+ \cos K_z\bigr)
+2t' \cos (K_y+K_z)
\eqno(1)
$$
where $t'/t\approx 0.8$ and 
$K_y,K_z$ refer to the
new coordinates in the extended dimer model BZ                  
which are {\it rotated} by $\pi/4$
compared to the those of the original BZ (see Fig. 1c).
 
A system in which Coulomb correlations are 
screened to be short range (Hubbard type)
may generically show
electron-phonon scattering {\it dominated by forward 
processes} \cite{CorrSmallq,PRBbig}.
In that case the effective pairing potential
takes the following form in 
momentum space \cite{PRBbig}:
$
V({\bf k}, {\bf k'})=
-{V\over {\bf q_c}^2+({\bf k}-{\bf k'})^2} + \mu^*({\bf k}-{\bf k'})
$.
The pairing kernel is characterized by a smooth momentum 
cut-off ${\bf q_c}$ which selects the small wavevectors
in the attractive phonon part while at larger wavevectors the
repulsive Coulomb pseudopotential $\mu^*({\bf k}-{\bf k'})$ may prevail.
This type of potential has been considered
for high-$T_c$
cuprates \cite{PRBbig,PRBrc,Abrikosov,Leggett,Weger}
and heavy fermion systems \cite{Gorkov}.
Screening by {\it short range} Hubbard-like Coulomb terms 
is necessary for obtaining such an effective small-q phonon pairing 
\cite{CorrSmallq,PRBbig} 
and thus we may naturally observe it in systems in which
the insulating phases show {\it AFM correlations} 
as in $\kappa-(ET)_2X$ salts.

Self-consistent solutions of the BCS gap equation     
with the small-{\bf q} pairing kernel and 
the electronic dispersion
of the dimer model are obtained using a fast Fourier
transform technique. The extended BZ (Fig. 1c)
has been discretized                 
with a $256 \times 256$ momentum grid.
Our d-wave solutions (see Fig. 2a) have
an {\it X-shape} remarkably similar 
with the experimental one \cite{Schrama}.
The angular position of the nodes and maxima
in the calculated and measured 
order parameters are in full agreement 
(experimental notations refer to the rotated coordinates of
the original BZ)
corroborating both
the relevance of the experimental technique employed in Ref.
\cite{Schrama} and that of our picture in 
$\kappa-(ET)_2X$'s.
Our d-wave solutions 
are in competition with
anisotropic s-wave solutions like the one shown in 
Fig. 2b. With a local Coulomb pseudopotential $\mu^*/V\approx 0.1$,   
for $q_c<\pi/4.5$ the d-wave solution prevails while for
$q_c>\pi/4.5$ the gap is anisotropic s-wave.

Engineering a physical situation in which
$t'/t\approx 1$, possibly by applying uniaxial
pressure, may be useful                       
for distinguishing which one of the
small-{\bf q} or the SF pairing pictures is relevant.
In our scheme the $t'/t=1$ system is still
a SC showing a multi-peak structure in the
momentum structure of the gap which could
perhaps be observable in an experiment
like in Ref. \cite{Schrama}.
We show in Fig. 2c
a typical d-wave SC gap solution obtained with our kernel
and $t'/t=1$.
Moreover, enhancing $t'/t$ within our
scheme advantages the d-wave solutions
in their close competition with anisotropic s-wave.
In a spin fluctuations mechanism instead, when $t'/t=1$ 
{\it there are no SC correlations} for $U/t<16$
\cite{Moriya} while for $U/t >> 1$ the problem
is mapped to a Heisenberg model on a regular
triangular lattice which is a well studied
frustrated system \cite{Anderson}. 

When small-{\bf q} processes dominate
the pairing, we have the situation of {\it Momentum Decoupling}
(MD) in SC meaning a tendancy for decorrelation 
of the SC behavior in 
the various regions of the FS \cite{PRBbig,PRBrc}.
This {\it loss of rigidity} of the momentum structure of the gap
leads to a distinct SC behavior exhibiting
density of states
driven anisotropies and a gradual {\it marginalization} of the 
SC gap symmetry for the condensation
free energy \cite{SSC,PRBbig}. 
The position of the gap maxima at the intersection of the
dimer FS with the original BZ in Fig. 2a, the anisotropic character
of the s-wave solution in Fig. 2b and the multipeak structures in
Fig. 2c reflect corresponding anisotropies of the density of states.
Doping induced variations
of the effective Coulomb pseudopotential
and other details in the pairing kernel \cite{PRBbig,SSC}
or variations in the concentration of impurities
or disorder \cite{Abrikosov}  
have been shown to induce
transitions between anisotropic s and d-wave SC.
The conflicting reports about the presence of nodes
in $\kappa-(ET)_2X$ SC are probably singatures    
of the {\it momenum softness} of the
SC gap and/or of the resulting marginality of the gap symmetry.
In the case of SF pairing the gap structure is
instead {\it rigid} and all experiments should report   
a d-wave gap.     

We show below that in the MD regime 
an {\it in-plane magnetic field} 
involved for example in NMR experiments        
may also influence significantly the
momentum shape of the gap and possibly induce {\it gap symmetry 
transitions} as well.
In the presence of a Zeeman field $H_Z$
we solve the BCS gap equation
$$
\Delta_{\bf k}=
\sum_{\bf k'}
{V_{\bf k,k'}\over \sqrt{\xi_{\bf k'}+\Delta_{\bf k'}}}
\biggl(
\tanh {\sqrt{\xi_{\bf k'}+\Delta_{\bf k'}}+H_Z
\over 2 T}
$$
$$
+\tanh {\sqrt{\xi_{\bf k'}+\Delta_{\bf k'}}-H_Z
\over 2 T}\biggr)
\eqno(2)
$$
This equation accounts for the Pauli effects
on SC and is particularly relevant when the field is
applied {\it parallel to the conducting planes} in which
case orbital effects are negligible.
We illustrate the effect of the field 
using a simplified
two dimensional square lattice model with nearest neighbors
hopping $\xi_{\bf k}=t (\cos k_xa + \cos k_ya)$.

We show in Fig. 3 the evolution of the d-wave gap
with the applid field along the first quarter of the
FS (connecting the $(\pi, 0)$ and $(0,\pi)$ points). 
For fields larger than about $H_c/3$, the 
shape near the nodes is modified significantly.
Approaching $H_c$ ($H_Z>0.8Hc$) we observe {\it extended
effectively
gapless regions around the nodes} whose extension
grows with the applied field. This behavior may have
significant experimental consequencies. For example,
the T-exponent of the penetration depth 
at low-T depends on the gap shape near the node.
Moreover,
the coexistence of extended gapless regions with SC regions
on the FS is the essential characteristic of
the Fulde-Ferrel-Larkin-Ovchinikov
state \cite{FFLO} and recent experiments claim the
observation of signatures of this state in
$\kappa-(ET)_2Cu(NCS)_2$ \cite{FFLOexp}.
An inhomogeneous SC state like the one shown in Fig. 3 near $H_c$
may possibly be at the origin of the
observations in Ref. \cite{FFLOexp}.

Furthermore,
we may obtain gap symmetry transitions
induced by the magnetic field. We show in 
the inset of Fig. 3
the evolution of the condensation free energy as a function
of the field in two characteristic cases of $\mu^*/V$
when $q_c=\pi/6$.
Enhancing the field we may obtain at low-T   
transitions
from s-wave to d-wave ($\mu^*/V=0.053$) or even {\it reentrant} transitions 
from d-wave to s-wave and then back to 
d-wave SC ($\mu^*/V=0.057$). A detailled study of
the field effects will be  
reported  
elsewhere, it is however plausible that
such field-induced s-d transitions may be responsible
for the quasi-systematic reports of d-wave SC
from experiments like NMR involving large {\it in plane} fields
while specific heat measurements in the absence of a 
field usually report a nodeless state.

In conclusion, we have shown that phonon mediated pairing
dominated by forward processes reproduces accurately the angular dependence
of the gap in $\kappa-(ET)_2X$ salts reported recently.
We argue that the experimental
conficts about the gap symmetry may be simple consequencies of the
{\it softness} of the momentum shape of the gap and the possible
near degeneracy of s- and d-wave SC in our picture.
We show that 
magnetic fields applied parallel to the planes
may induce extended gapless regions
accounting for recent indications of a FFLO
state and may plausibly induce
gap symmetry transitions as well.      

I am indepted to Michael Lang for numerous illuminating discussions
about the experiments.

\newpage

\begin{figure}[tbp]
\caption{
a) The $\kappa$ packing motif. Each stick corresponds to
a BEDT-TTF molecule. b) The effective frustrated
lattice scheme in the dimer model approximation.
c) The FS (thick line) of the dimer model in the extended
BZ scheme. The original BZ is shown with dotted lines.
In the real system,
there is a small gap openning
at the intersection of the dimer model FS with the original BZ
leading to a hole-like
FS sheet around the Z point
and a quasi-1D sheet along the z-axis}
\end{figure}

\begin{figure}[tbp]
\caption{
Self-consistent gap solutions with small-q pairing over the extended
BZ of the dimer model: a) typical d-wave solution, b) 
competing anisotropic s-wave solution and c) d-wave solution obtained
when $t'/t=1$.}
\end{figure}

\begin{figure}[tbp]
\caption{
Typical evolution of the d-wave gap in the small-q pairing scheme
($q_c\approx \pi/10$) over
the first quadrant of the BZ of a NN hoping square
lattice model
with $H_Z=0$ (full line), $H_Z=0.75 H_c$ (dotted line)
$H_Z=0.9 H_c$ (dashed line) and $H_Z=0.975 H_c$
(dotted-dashed line).
In the inset is shown the
evolution of the absolute condensation free-energy F
as a function of the field H applied parallel to the planes
(both in arbitrary units)
for the d-wave (full line) and the s-wave states
with $\mu^*/V=0.53$ (dotted line)
and $\mu^*/V=0.57$ (dashed line)
when $q_c=\pi/6$
(the d-wave state is insensitive to a local $\mu^*$).}
\end{figure}

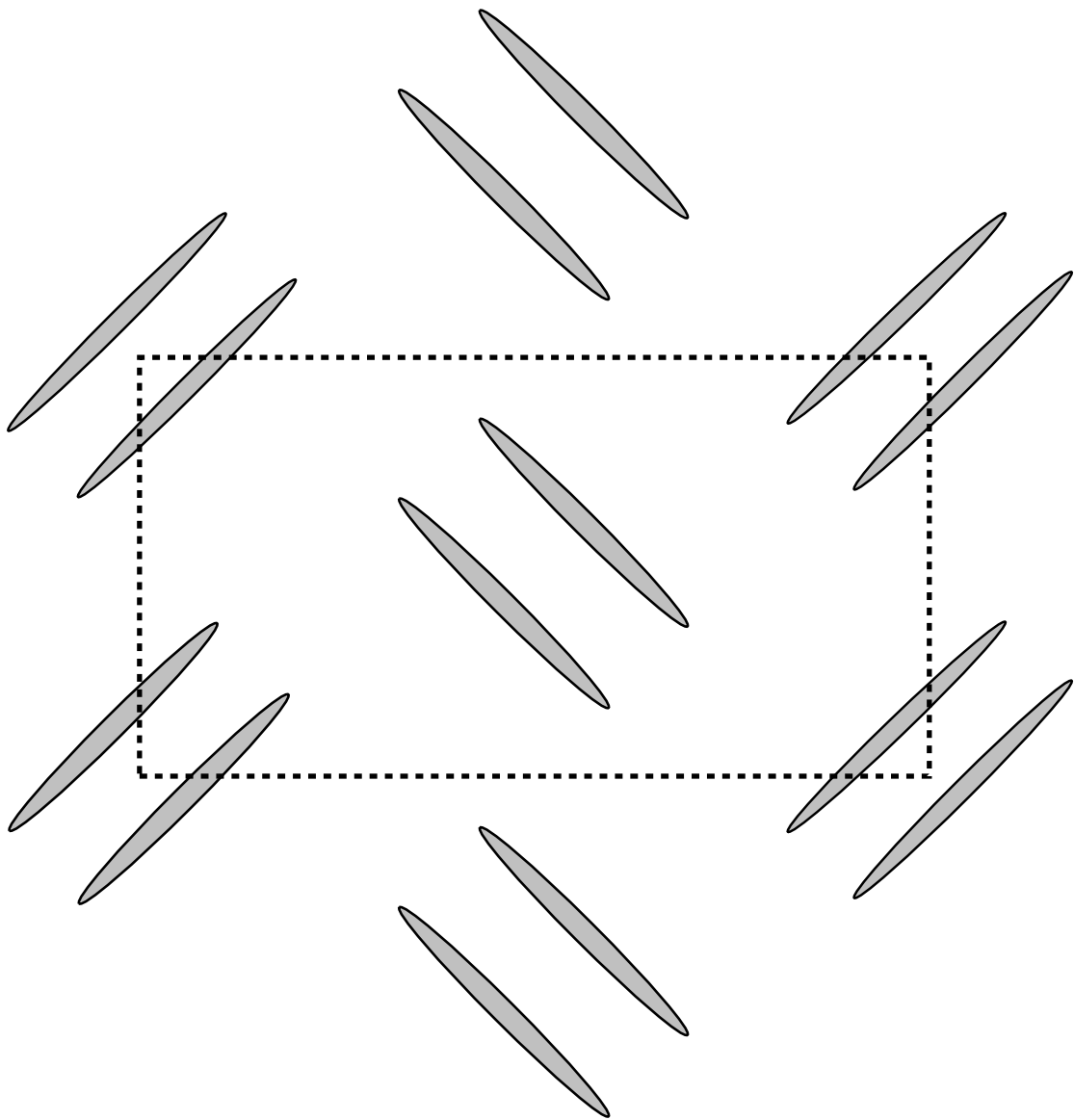
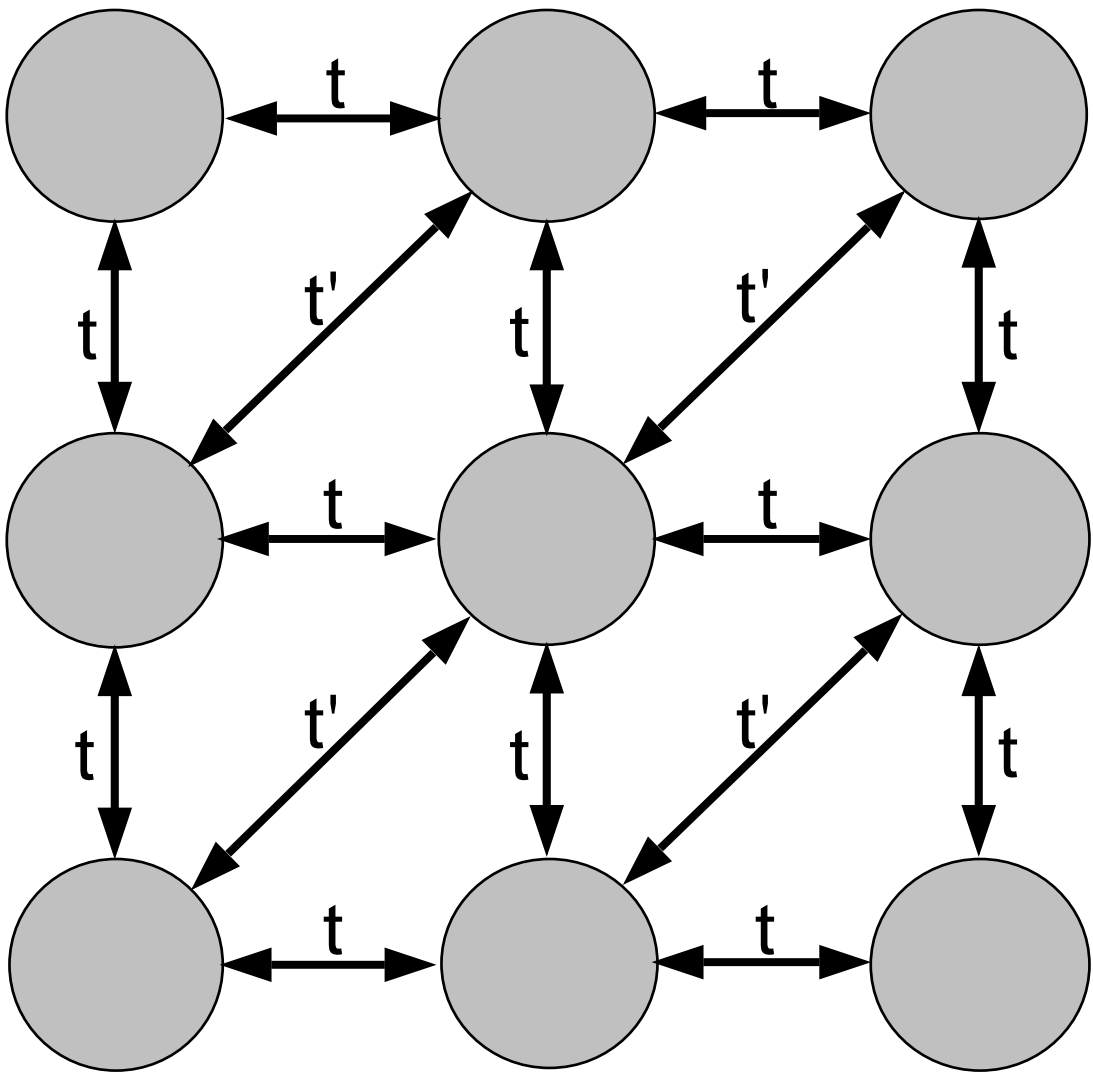
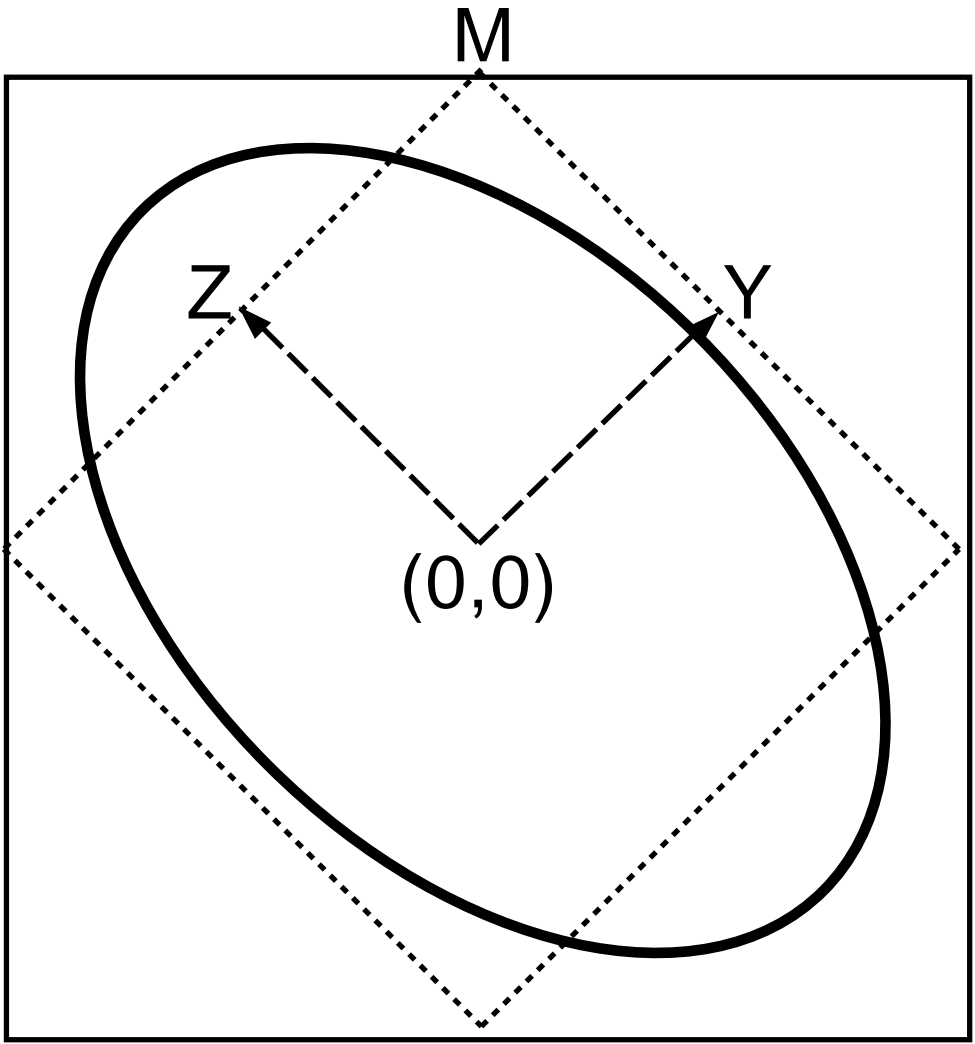
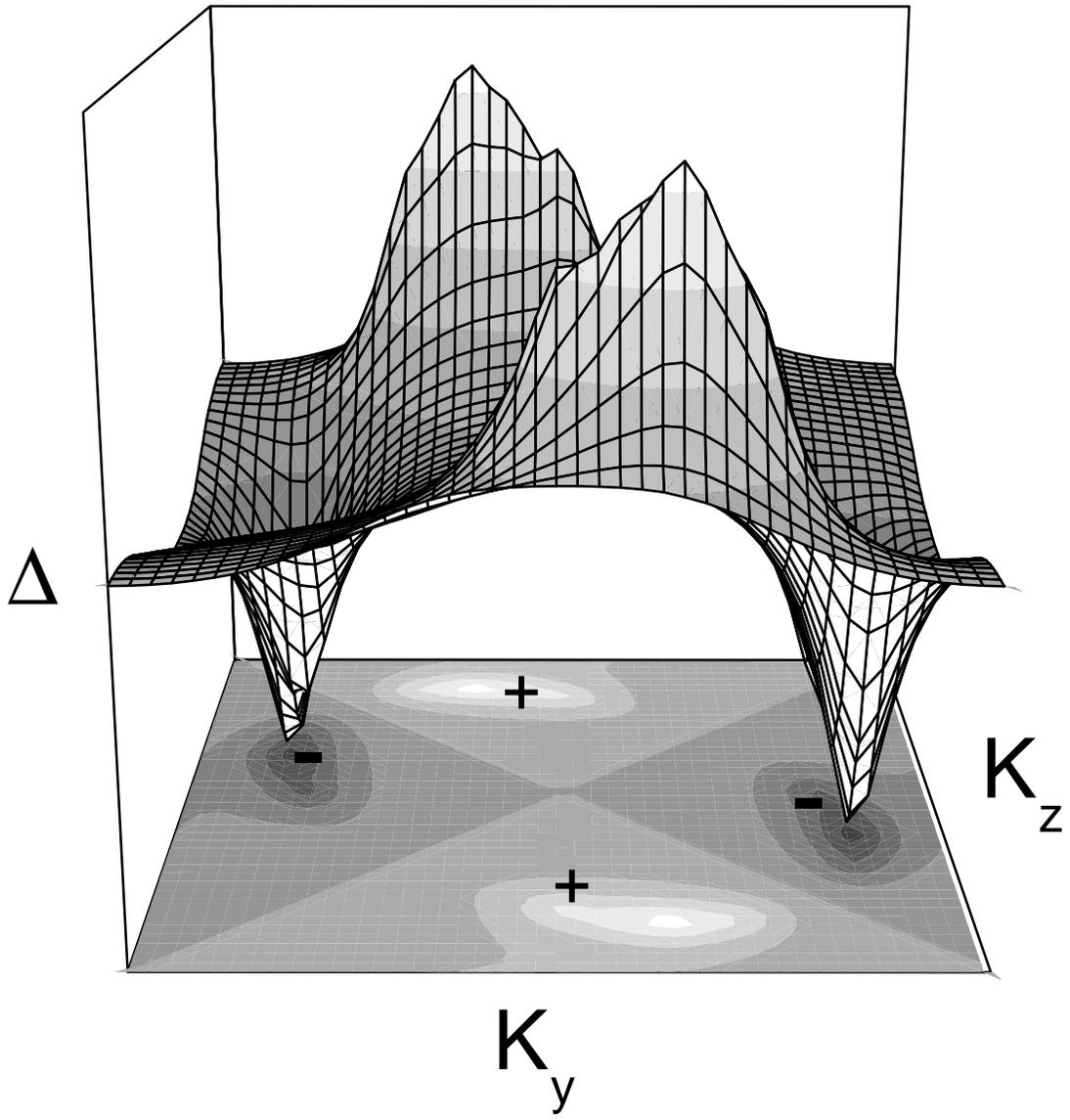
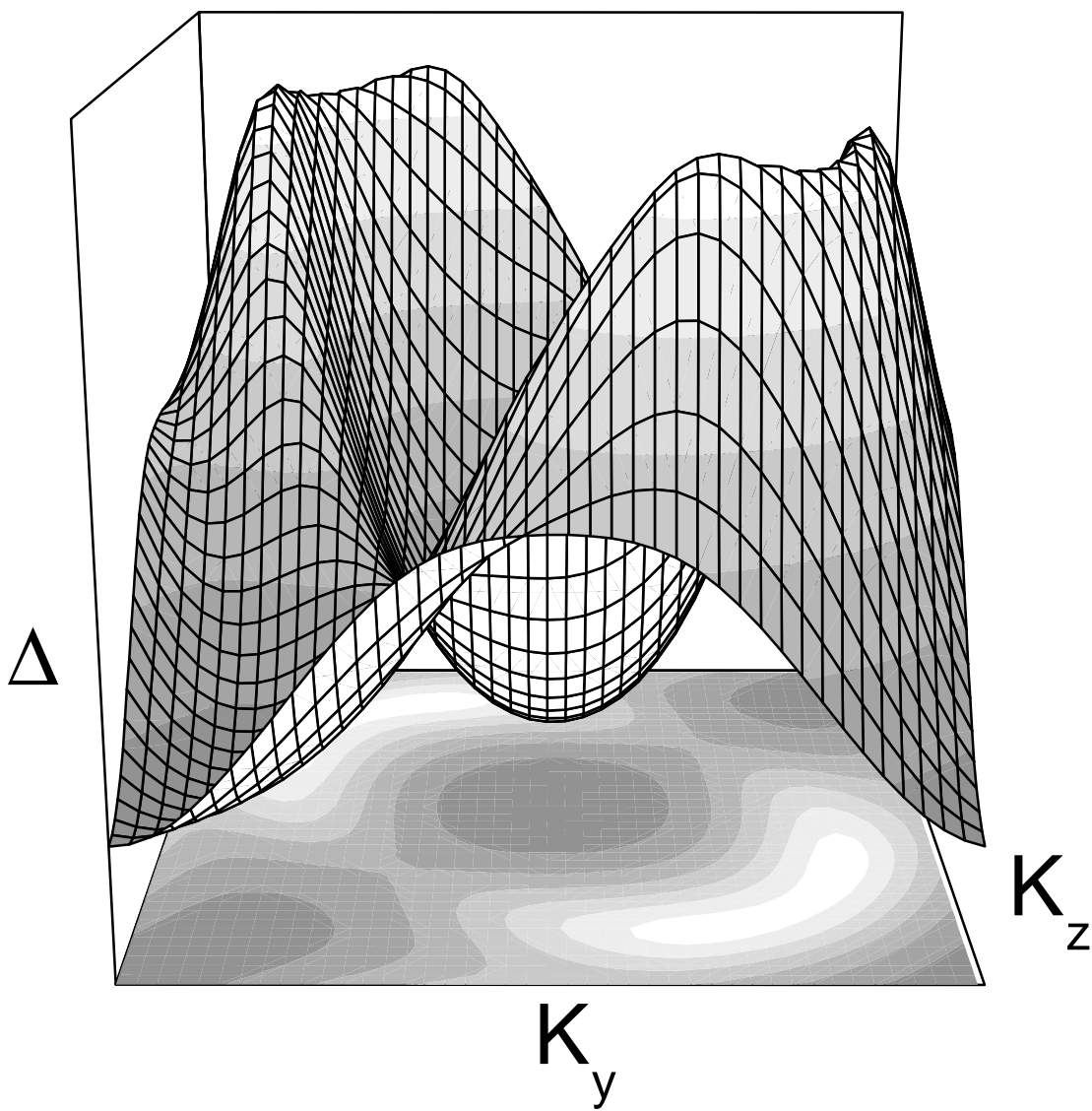
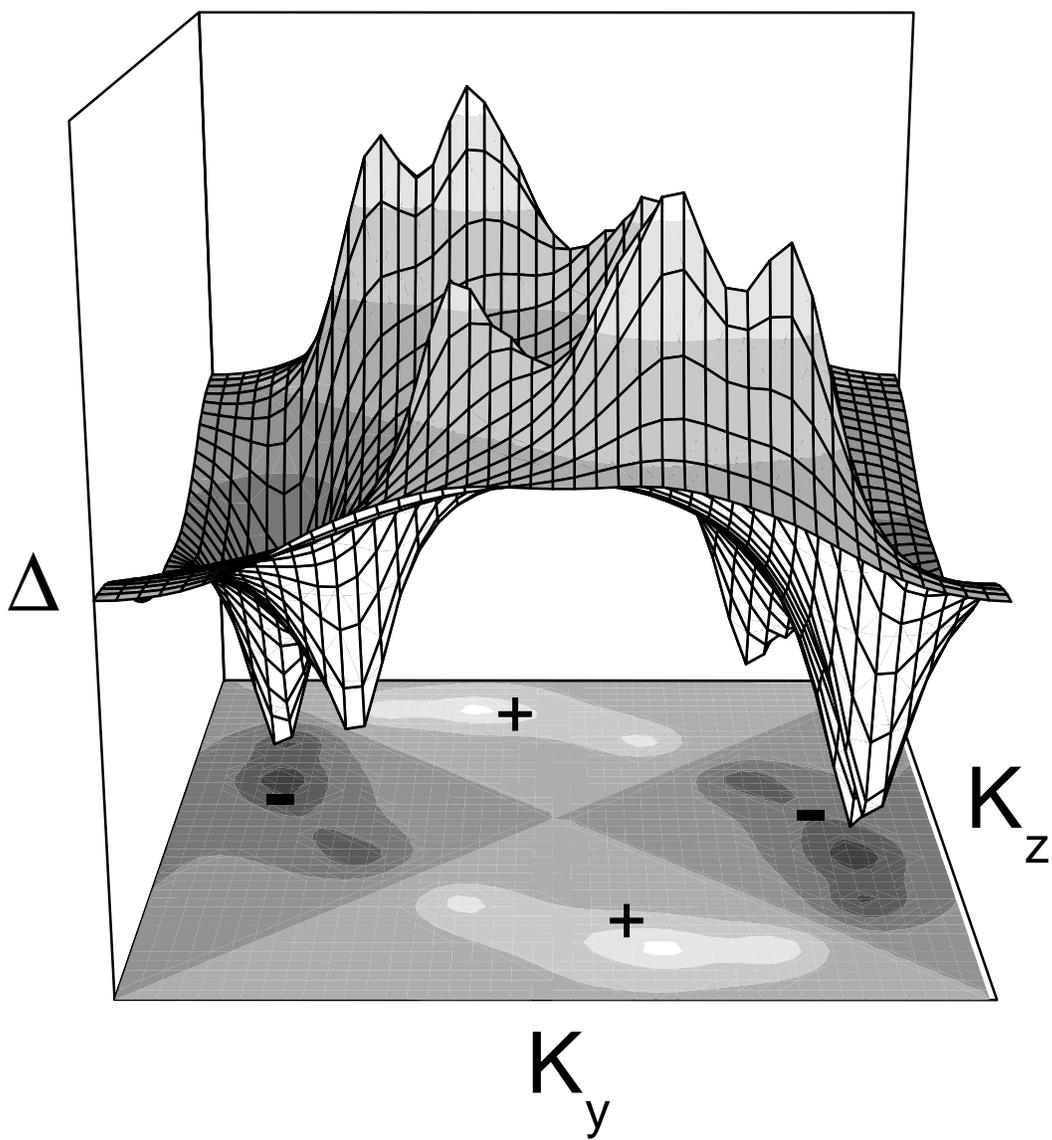
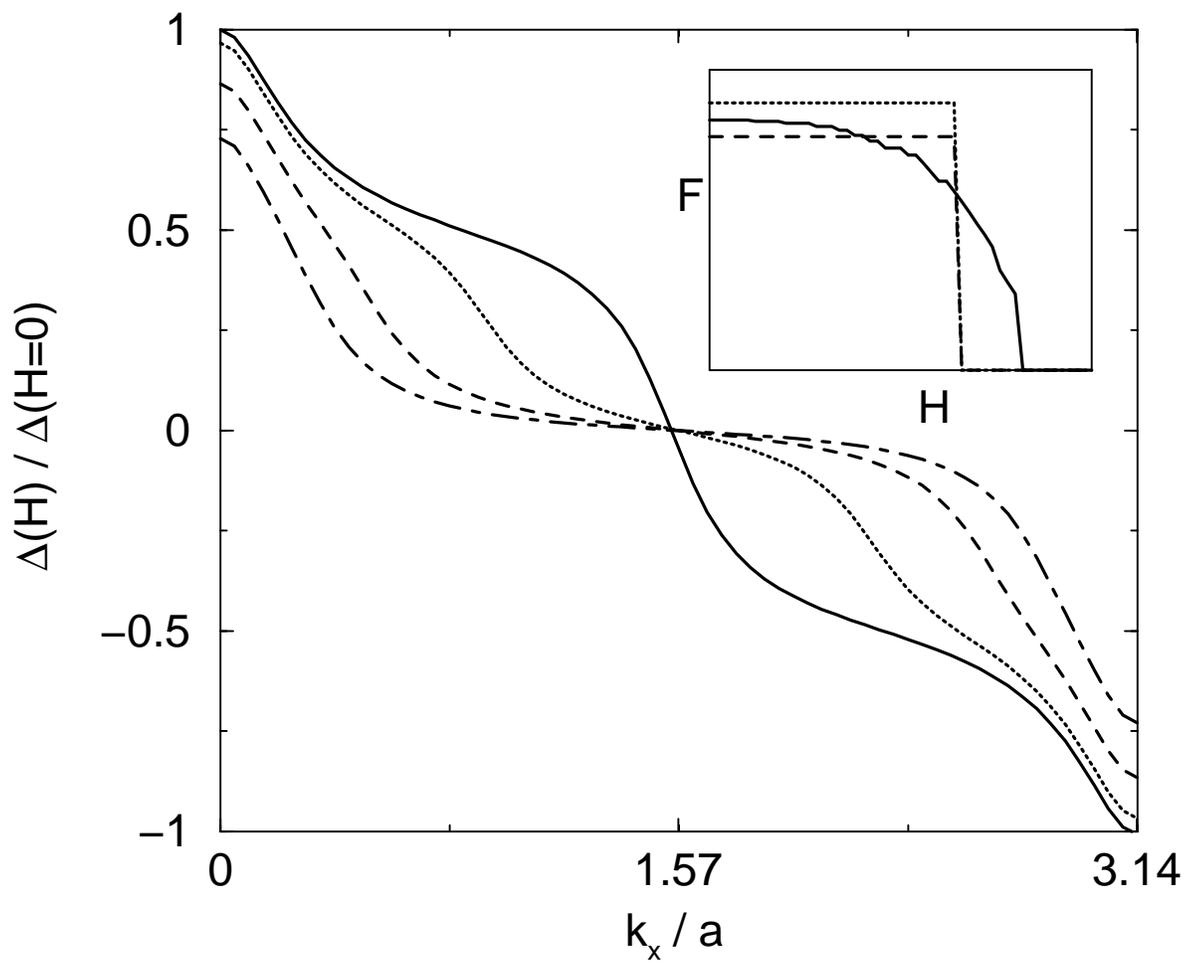
\end{document}